\documentstyle[prb,aps]{revtex}
\begin{document}
\draft
\title{Gate-controlled spin polarized current in ferromagnetic
single electron transistors}
\author{Watson Kuo$^{1,2}$\cite{email} and C. D. Chen$^1$}
\address{ $^1$Institute of Physics, Academia Sinica,
Nankang, 115, Taipei, Taiwan, ROC \\ $^2$Department of Physics,
National Tsing Hua University, Hsinchu, 300, Taiwan, ROC \\}
\date{September 21, 2001}
\maketitle

\begin{abstract}
We theoretically investigate spin dependent transport in
ferromagnetic/normal metal/ferromagnetic single electron
transistors by applying master equation calculations using a two
dimensional space of states involving spin and charge degrees of
freedom. When the magnetizations of ferromagnetic leads are in
anti-parallel alignment, the spins accumulate in the island and a
difference of chemical potentials of the two spins is built up.
This shift in chemical potential acts as charge offset in the
island and alternates the gate dependence of spin current. Taking
advantage of this effect, one can control the polarization of
current up to the polarization of lead by tuning gate voltages.
\end{abstract}

\pacs{PACS numbers: 72.25.-b, 73.23.Hk, 85.75.-d, 85.35.Gv}

Ferromagnetic single electron transistor (SET) has been an
interesting system shown to exhibit novel phenomena with an
interplay between spin and charge. Recently, Chen {\it et al}
\cite{ccd} and Ono {\it et al} \cite{ono} succeeded in fabricating
small double junctions containing metal or superconductors
sandwiched between two ferromagnets. In their experiments,
enhanced tunneling magnetoresistance(TMR), magneto Coulomb
oscillations and spin accumulations were observed. On the other
hand, theories of ferromagnet/ferromagnet/ferromagnet(F/F/F) and
ferromagnet/normal metal/ferromagnet(F/N/F) SET's based on
transition rate and master equation formalism were developed to
derive bias-voltage and gate-voltage dependent TMR in both
sequential and strong tunneling
regimes.\cite{taka1,barnas,majumdar,koro,brataas,wang} The pioneer
experiment conducted by Johnson and Silsbee demonstrated the
importance of spin accumulation effect on ferromagnet-normal metal
systems.\cite{mark} For F/N/F double junctions, the spin
accumulation is predicted to occur when two ferromagnet leads are
in anti-parallel alignment, which would lead to a new origin of
TMR in contrast to that of F/F/F cases.\cite{taka1,koro,brataas}
In this study, we investigate the spin accumulation and related
phenomena in F/N/F SET under the influence of gate charge.

The spin dependent transport in a ferromagnet is usually described
by the relative difference of the majority and minority spins of
conduction electrons, denoted as polarization $P$.\cite{review}
Under the condition that spins do not flip, the transport current
in F/N/F double tunnel junctions can be separated into two
channels labeled as upspin and downspin which, throughout this
article, are assumed to be contributed by majority and minority
spins, respectively, in the source ferromagnet. For example, when
lead magnetizations are in anti-parallel alignment, the upspin
channel has a larger tunneling rate for the source junction than
for the drain junction. In this case and in the steady condition,
the upspin chemical potential in central electrode rises to
balance the spin's incoming and outgoing rates, and the chemical
potential of the downspin would decrease by the same amount. This
shift in spin chemical potential, denoted as
$\Delta\mu_{\uparrow(\downarrow)}$, is predicted to be $\pm
PeV_b/2$, in which $V_b$ is the applied bias voltage (see inset of
Fig. 1) and $P$ is the polarization of the two leads. Therefore,
the net spin in normal metal becomes nonzero and this effect is
known as the spin accumulation (or spin imbalance).

A gate voltage $V_g$ is applied to turn on and off the charge
transport in SET by tuning the electrostatic potential of the
island. When SET is symmetrically biased with $V_g$=0, an energy
cost of roughly the charging energy $E_C$ is required for adding
or removing an excess charge in the island, and the charge
transport is blockaded provided that temperature is low, $k_BT\ll
E_C$. When the gate voltage is tuned with two adjacent charge
states becoming energetically degenerate, electrons can enter or
leave the island without extra energy cost, producing sequential
tunneling current. Ideally, the current is at a minimum and
maximum, respectively, for $V_g=0$ and $e/2C_g$ and can be
modulated periodically with a period of $\Delta V_g=e/C_g$; here
$C_g$ is the island-to-gate capacitance. These two gate voltages
are thus referred to as minimal and maximal gate voltages. Our
study suggests that the shift in spin chemical potentials
generated by spin accumulation produces an effective charge
offsets.

To demonstrate this charge offset effect, we apply a modified
master equation calculation which takes into account the spin
dependent charge states of the island. In this framework, the
states are described by two parameters, $Q_\uparrow$ and
$Q_\downarrow$, denoting excess upspin charge and excess downspin
charge, respectively. Comparing with spin-independent case, each
primitive charge state $|Q\rangle$ is built up by spin charge
states of $Q=Q_\uparrow+Q_\downarrow$. Generally speaking, these
spin dependent charge states are in non-equilibrium with the
presence of spin accumulation. However, under limit of short
energy relaxation time, the occupation distribution of a
particular spin would form an equilibrium Fermi distribution,
$i.e.$ $f(\varepsilon-\mu_{\uparrow(\downarrow)})$. The numbers of
net spin $N=(Q_\uparrow-Q_\downarrow)/e$ is related to the spin
chemical potentials and the density of states of the island
$\rho(\varepsilon)$ as $ N=\int d\varepsilon \rho(\varepsilon)
\left(f(\varepsilon-\mu_{\uparrow})-f(\varepsilon-\mu_\downarrow)\right).$

Although the electrostatic energy of each charge state is spin
independent, the tunneling rate is spin dependent in two ways:
first, the effective tunneling resistances for the majority and
minority spin tunneling processes are multiplied respectively by
$2/(1-P)$ and $2/(1+P)$; second, the spin chemical potential
shift's presence also modifies transition rates by changing the
numbers of possible tunneling processes. Consequently, the master
equation for each spin charge state, together with certain spin
flipping transitions, reads
\begin{eqnarray}
\frac{dp_{ij}}{dt}&=&\sum_{l={\rm
S,D}}\left\{\Gamma_\uparrow^l(i,j|i\pm1,j)p_{i\pm1,j}+
\Gamma_\downarrow^l(i,j|i,j\pm1)p_{i,j\pm1}\right\} \nonumber \\ &
&-\sum_{l={\rm S,D}}\left\{\Gamma_\uparrow^l(i\pm1,j|i,j)+
\Gamma_\downarrow^l(i,j\pm1|i,j)\right\}p_{ij}+\left(\frac{dp_{ij}}{dt}\right)_{\rm
sf }, \label{master}
\end{eqnarray}
in which $i=Q_\uparrow/e$ and $j=Q_\downarrow/e$ denote the
numbers of upspins and downspins respectively, and
$\Gamma_s^l(i^\prime,j^\prime|i,j)$ is the tunneling rate for spin
direction $s$(=$\uparrow$, $\downarrow$) in junction $l$ (S for
source and D for drain) from states $|i,j\rangle$ to
$|i^\prime,j^\prime\rangle$, and $p_{ij}$ is the probability that
the island is in state $|i,j\rangle$. In this equation, only
sequential tunneling process is considered. By introducing an
energy-independent spin relaxation time $\tau_s$, the spin
flipping transitions can be explicitly written as:
\begin{equation}
\left(\frac{dp_{ij}}{dt}\right)_{\rm sf
}=\frac{1}{\tau_s}\left\{\frac{ U /\delta}{1-\exp{(-\beta U )}}
p_{i+1,j-1}+\frac{- U /\delta}{1-\exp{(\beta U )}}p_{i-1,j+1}
\right\}-\frac{1}{\tau_s}\frac{ U }{\delta}\frac{1+\exp{(-\beta U
)}} {1-\exp{(-\beta U )}}p_{ij}
\end{equation}
The first and second terms describe respectively the increase of
probability due to up-to-down and down-to-up flip processes, and
the third term is the decrease arising from the opposite
processes. Here we assume that the up-to-down spin flipping rate
is proportional to
$f(\varepsilon-\mu_\uparrow)(1-f(\varepsilon-\mu_\downarrow))$ for
electron with energy $\varepsilon$. $U
=\mu_\uparrow-\mu_\downarrow=(i-j)\delta$ is the chemical
potential difference of upspins and downspins, and $\delta\sim
1/\rho$ is the energy level spacing, a constant for normal metal.
For positive $ U $, up-to-down spin flip is favorable, while for
negative $ U $, down-to-up spin flip dominates. Under these
conditions the probabilities of major spin states with large $| U
|$ are greatly reduced while suppressing the spin accumulation.

Eq.\ (\ref{master}) can be solved under the stationary condition
given by $dp_{ij}/dt=0$ as described in the spin independent
case.\cite{master} Through a particular distribution of $p_{ij}$,
one can obtain the amount of spin accumulation, quantified as
average chemical potential difference,
$\overline{U}=\sum_{i,j}(i-j)\delta p_{ij}$, and the spin current
for spin $s$ tunneling through junction $l$,
\begin{equation} I_{s}^{l}=e\sum_{ij} \left[
\Gamma_s^{l}(i+1,j|i,j) -\Gamma_s^{l}(i-1,j|i,j)\right] p_{ij}.
\end{equation}
If there is no spin flipping processes, the spin is conserved and
the spin current passing through the source and drain junctions is
the same. If the spin flips too quickly so as to completely
destroy the spin accumulation, then the ratios between the two
spin currents, $I_{\uparrow}/I_{\downarrow}$, for source and drain
junctions, will be the same as polarization of source and drain
electrodes, respectively.

To gain an understanding about this phenomena, here, we perform a
simulation using device parameters similar to those in experiments
{\it et al} \cite{ccd}: $R_{\rm S}=R_{\rm D}$ =400k$\Omega$,
$C_{\rm S}=C_{\rm D}$=300aF, $C_g$=0.8aF, $P_{\rm S}=P_{\rm
D}$=0.4. Because the resistances are much higher than quantum
resistance $R_Q$, the contribution due to higher order tunneling
processes is negligible and only sequential tunneling process is
included. We consider $IV_b$ characteristics and current-gate
voltage dependences ($IV_g$) with both parallel and anti-parallel
alignment of leads under the no spin-flipping condition at a
temperature of $k_BT/E_C=0.1$. In the parallel configuration, no
particular feature is found because the ratio between two spin
currents is simply the polarization $0.4$, and total current is
the same as that of the spin independent case. When the leads are
in anti-parallel alignment, the calculation provides much more
interesting results. The total current is smaller than that of the
parallel case, and the high bias differential resistance is
increased by a factor of $1/(1-P^2)$ and shows a generic F/N/F TMR
effect. The differential TMR as a function of bias voltage also
exemplifies expected oscillatory behaviors.\cite{brataas}

The $IV_g$ characteristics shown in Fig.\ \ref{ivg}a exhibit
particularly different behaviors than from the parallel case. The
upspin and downspin currents are only the same at $V_g=0$ and
$V_g=e/2C_g$. A closer inspection reveals that the peaks of two
$IV_g$ curves with opposite spins shift with increasing bias
voltage. This effect can be explained when we consider two
separated spin transport channels. When the leads are in
anti-parallel alignment, the source and drain resistances for a
particular spin channel may differ by several times. This results
in a step-like structure, called Coulomb staircase, in the $IV_b$
characteristics, and a distorted saw-tooth-like $IV_g$ modulation.
The Coulomb staircase effect can explain the TMR oscillation and
the asymmetric gate dependence of spin
currents.\cite{barnas,majumdar} However, for a more rigorous
study, it is necessary to include the spin entanglement term in
the Hamiltonian, $2E_CQ_\uparrow Q_\downarrow/e^2$. In fact, our
calculations suggest that the results of the two methods differ
especially at high bias voltages where both $Q_\uparrow$ and
$Q_\downarrow$ are large.

From the view point of spin accumulation, the raised upspin
(lowered downspin) chemical potential effectively gives rise to a
positive (negative) charge offset. At low bias voltage regime
($V_b<2E_C/e$), when $V_g$ is gradually raised from zero to
maximal value(=$e/2C_g$), the electrical current increases due to
suppression of Coulomb blockade. The spin accumulation is, in
turn, enhanced by the increased current, and consequently there is
a rise in both upspin chemical potential and the effective charge
offset. Since within $0<V_g<e/2C_g$ region, the charge offset is
an ascending function of $V_g$, and the upspin current increases
more rapidly than that of the zero spin accumulation. On the other
hand, the increment of downspin current is less effective, because
the downspin chemical potential decreases as $V_g$ increases.
Therefore, at low bias regime, the $IV_g$ for upspin and downspin
are tilted respectively toward lower and higher $V_g$ directions,
and form saw-tooth-like $IV_g$ patterns. At bias voltages far
beyond threshold ($V_b\gg2E_C/e$), the current is not much
affected by $V_g$ and the spin chemical potential is less
sensitive to $V_g$. The shift in spin chemical potential (relative
to the no spin accumulation case)
$\Delta\mu_{\uparrow(\downarrow)}$ increases (decreases) with
$V_b$. At $V_b=6E_C/e$, $\Delta\mu_{\uparrow(\downarrow)}$ is
about $\pm E_C$, corresponding to a charge offset of about $\pm
e$. Consequently, as shown in Fig.\ \ref{ivg}a, the upspin and
downspin $IV_g$ characteristics shift by one period in respect
with each other and differ from $IV_g$ characteristics at low bias
voltages by half period. The total current is shown in Fig.\
\ref{ivg}b, allowing a comparison with the experiments.

For further investigation of the effect of applied gate voltage on
two spin currents, we define a quantity describing the
polarization of the tunneling current:
$P_I=(I_\uparrow-I_\downarrow)/(I_\uparrow+I_\downarrow)$. Fig.\
\ref{poi} shows bias and gate voltage dependence of $P_I$. Such
dependence suggests the possibility of using a ferromagnetic SET
as a gate-controlled current polarizer. Because of large $P_I$
values, the optimum operating regime is at low bias voltage. At
$eV_g/C_g=\pm 0.15$ and $eV_b/E_C=0.7$, the current polarizations
reach a maximum value of $\pm0.33$. One can also explore the
temperature dependence of the polarization current. There are two
ways that the effects of temperature can enter, both leading to
the destruction of current polarization. One is thermal activated
charge fluctuation and the other is decrease of spin flip time.
The former is automatically included in the master equation
calculation and its effect is shown in Fig\ \ref{poits}(a). At
$T=0$, the value of $P_I$ can be as large as the polarization of
the lead itself, while at $k_BT\gtrsim 0.5E_C$, the gate charge
effect becomes negligible. To evaluate the effect of the spin flip
process, we assume an energy independent spin flipping time and an
energy level spacing $\delta$ of $1\mu e$V in the island, and
perform the calculations using the same device parameters as
above. Fig\ \ref{poits}(b) shows the current polarization for
drain junction operating at $V_b=0.7E_C/e$ as a function of gate
voltage at $k_BT/E_C=0.1$ under several spin flipping times.
Clearly, when the spin flip time is short as compared with the
tunneling time $\tau_t=e/I$ of approximately $10$ns, the spin
accumulation diminishes and $P_I=-0.4$, which is simply the
polarization of the drain electrode. However, since the chemical
potential is proportional to the island's density of states, the
required spin flip time would be shorter for nanometer-sized
normal-metal islands in which the level spacing is of the order of
$10^{-8}\sim 10^{-9} e$V, which is much smaller than the assumed
value. It has been proposed that the criteria for spin
accumulation is related to the tunneling resistance $R_t$ as
$\tau_s\delta/\hbar>R_t/R_Q$.\cite{brataas} Our calculation
results agree with this prediction.

The co-tunneling processes, which are thus far not included in our
calculations, can also give induce effective spin flipping. In the
spin independent case, co-tunneling is a second order process that
preserves the charge state but also produces current. In the
Coulomb blockade regime, where sequential tunneling is suppressed,
the current is mainly due to co-tunneling. For spin co-tunneling,
there are spin-conserved and spin non-conserved processes. The
latter is that a spin enters the island and an opposite spin
leaves. The forward and backward co-tunneling rates
$\overrightarrow{\Gamma_{\rm co}}$, $\overleftarrow{\Gamma_{\rm
co}}$ for F/N/F SET can be written as\cite{cot}:

\begin{equation}
\overrightarrow{\overleftarrow{\Gamma_{\rm
co}}}=\frac{R_Q}{4\pi^2e^2 R_{\rm S,eff}R_{\rm D,eff
}}\int{d\varepsilon}\frac{\varepsilon}{1-\exp(-\beta\varepsilon)}
\frac{\Delta E-\varepsilon}{1-\exp[-\beta(\Delta
E-\varepsilon)]}\left|\overrightarrow{\overleftarrow{M}}\right|^2,
\label{coteq}
\end{equation}
where
\begin{equation}
\overrightarrow{\overleftarrow{M}}=\frac1{E_{\rm
S}^\pm+\varepsilon+i\gamma^\pm}+\frac1{E_{\rm D}^\mp+\Delta
E-\varepsilon+i\gamma^\mp}.
\end{equation}
The $\Delta E$ is the energy difference between initial and final
states. For spin conserved co-tunneling, $\Delta E=\pm eV$,
whereas for up-to-down and down-to-up co-tunneling $\Delta E=\pm
eV- U $ and $\Delta E=\pm eV+ U $, respectively (`+' for forward
and `-' for backward). $E_{\rm S}^\pm$ and $E_{\rm D}^\pm$ are
energy changes of the tunneling processes $Q\longrightarrow Q\pm
e$ for source and drain junctions. $R_{\rm S, eff}$ and $R_{\rm
D,eff}$ are the effective tunneling resistances for the source and
drain junctions. Note that for anti-parallel configuration,
$R_{\rm S,eff}R_{\rm D,eff}$ product in Eq.\ (\ref{coteq}) is
$R_{\rm S}R_{\rm D}$ product multiplied by $4/(1-P_{\rm
S})(1+P_{\rm D})$, $4/(1+P_{\rm S})(1-P_{\rm D})$, $4/(1+P_{\rm
S})(1+P_{\rm D})$ and $4/(1-P_{\rm S})(1-P_{\rm D})$ for up-to-up,
down-to-down, up-to-down and down-to-up forward co-tunneling
events, respectively. $\gamma^\pm$ are decay rates for the final
charge states $Q\pm e$ of the two processes and are given by
\begin{equation}
\gamma^\pm=\frac{R_Q}{4\pi}\left(\frac{E_{\rm S,eff}^\pm}{R_{\rm
S,eff}}\coth{\frac{\beta E_{\rm S,eff}^\pm}{2}}+\frac{E_{\rm
D,eff}^\pm}{R_{\rm D,eff}}\coth{\frac{\beta E_{\rm
D,eff}^\pm}{2}}\right).
\end{equation}
To investigate the co-tunneling spin flipping, we use tunneling
resistances of 40k$\Omega$, which is closer to $R_Q$, and leave
other parameters unchanged. In the Coulomb blockade regime, the
co-tunneling spin flipping rate can be as large as $10^8$ Hz,
which is comparable to the tunneling rate of $I/e\sim 10^9$Hz.
However, different from the spin relaxation as discussed above,
the spin flipping induced by co-tunneling for $| U |\ll eV_b$ does
not have preferred direction. That is, it does not smear out the
spin accumulation but rather only enhances spin fluctuation. Fig.\
\ref{cot} shows the spin fluctuation $\delta N=\sqrt{\overline{ U
^2}-\overline{ U }^2}/\delta$ as a function of gate voltage at
$V_b=0.75E_C/e$ with and without consideration of co-tunneling
processes.

In summary, we proposed theoretically a gate-controlled polarized
current in ferromagnetic single electron transistors. Using
modified master equation formalism, we calculate spin sequential
tunneling rates and spin current when spin accumulation is
present. The spin accumulation-induced chemical potential shift
behaves like charge offsets, producing interesting effects to the
$IV_g$ characteristics. When the gate voltage is tuned away from
the maximal and minimal gate voltages, the current passing through
the junctions is polarized. The thermal fluctuation and spin
flipping processes are both shown to suppress the effects from
charge offset. The co-tunneling event provides effective spin
flipping processes, but it only enhances spin fluctuation. Taking
advantage of the gate dependence of polarized current in a
ferromagnetic SET, one can utilize this system as a tunable
current polarizer.

We thank the members of the SET Group of the Institute of Physics,
Academia Sinica, Taiwan for fruitful discussions. This research
was partly funded by the Nation Science Council No.
89-2112-M-001-033.

\begin{figure}
\caption{Current-gate voltage dependences for (a) the spin
currents from $eV_b=0.5 E_C$(bottom) to 6.0$E_C$(top) and (b)
total current from $eV_b=0.5 E_C$(bottom) to 4.0$E_C$(top) with a
step 0.5$E_C$ for an F/N/F SET, whose parameters are described in
the text. In (a) the dotted and solid curves represent
respectively up-spin and down-spin currents. Notice that the peaks
and valleys for the two spin currents appear at different gate
voltages. At $eV_b=4.5 E_C$ the two currents shift half period
while near $eV_b=6.0 E_C$, they are the same. In (b) $IV_g$ curves
shift with bias voltages: at $eV_b=6.0 E_C$, it shifts by 0.5$e$.
(c) The scheme of the F/N/F SET considered. The arrows indicate
the magnetizations of electrodes.}\label{ivg}
\end{figure}

\begin{figure}
\caption{(a) Polarization of current dependences on bias voltage
at $eV_g/C_g$= 0.15(solid curve) and -0.25(dotted curve). At
$eV_b/E_C=0.7$ and $eV_g/C_g=\pm 0.15$, $P_I$ reaches a maximum
value of $\pm$0.33 while at $eV_b/E_C=4.0$ and $eV_g/C_g=\pm
0.25$, $P_I$ has another maximum, which is approximately 0.04 with
an opposite direction. (b) Polarization as a function of gate
voltage at $eV_b/E_C$=0.5 to 4.0 with a step 0.5. } \label{poi}
\end{figure}

\begin{figure}
\caption{(a) Gate voltage dependences of polarization at
$eV_b/E_C$=0.7 for temperatures ranging between $0$ and
$0.5E_C/k_B$ with a step 0.05 under no spin flip assumption. At
$T=0$, $P_I$ cannot be defined for $C_gV_g/e<0.33$ and
$C_gV_g/e>0.67$ since the current is zero within that range. The
effect is more pronounced at lower temperatures. (b) The same
dependence for the drain junction at $eV_b/E_C$=0.7 and
$k_BT/E_C=0.1$ with several spin flipping times $\tau_s$, from
bottom to top: 10ns, 100ns, 1$\mu$s, 10$\mu$s, 100$\mu$s (solid
curves), and $\infty$ (dotted curve). For $\tau_s=10$ns, $P_I$ is
fairly close to the $P$ value of the drain electrode, {\it i.e.}
no spin accumulation. For $\tau_s=100\mu$s, spin accumulation is
almost the same as the non-flipping case and $P_I$ increases
dramatically.} \label{poits}
\end{figure}

\begin{figure}
\caption{Spin fluctuation $\delta N$ v.s. gate voltage $V_g$ in
the Coulomb blockade regime, $V_b=0.1$mV. The solid and dotted
curves represent respectively the result for sequential tunneling
with and without co-tunneling processes. The fluctuation is
increased by about 20\%. The inset shows the average spin number
$N_{\rm ave}=\overline{U}/\delta$ for the two cases.} \label{cot}
\end{figure}

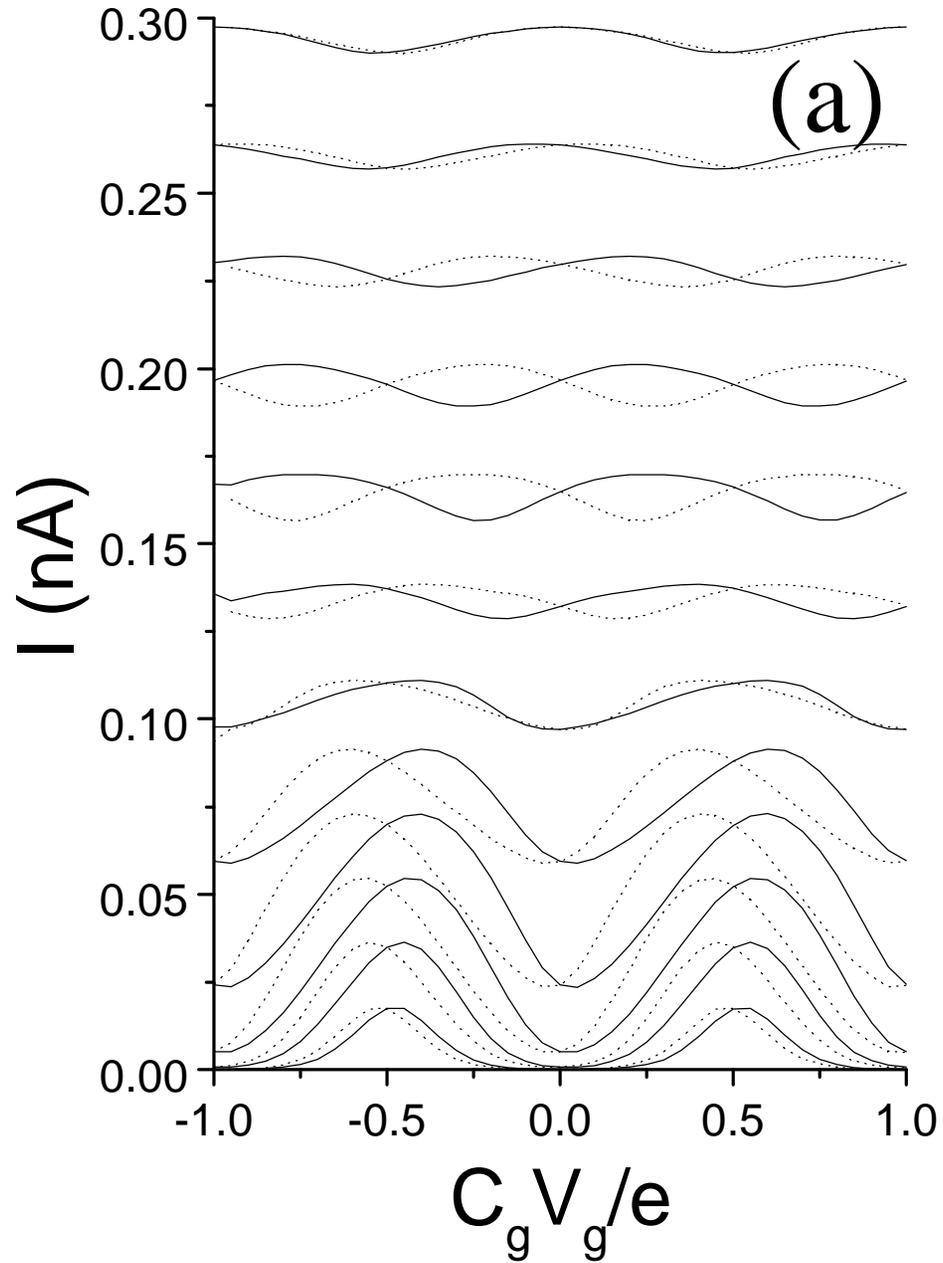
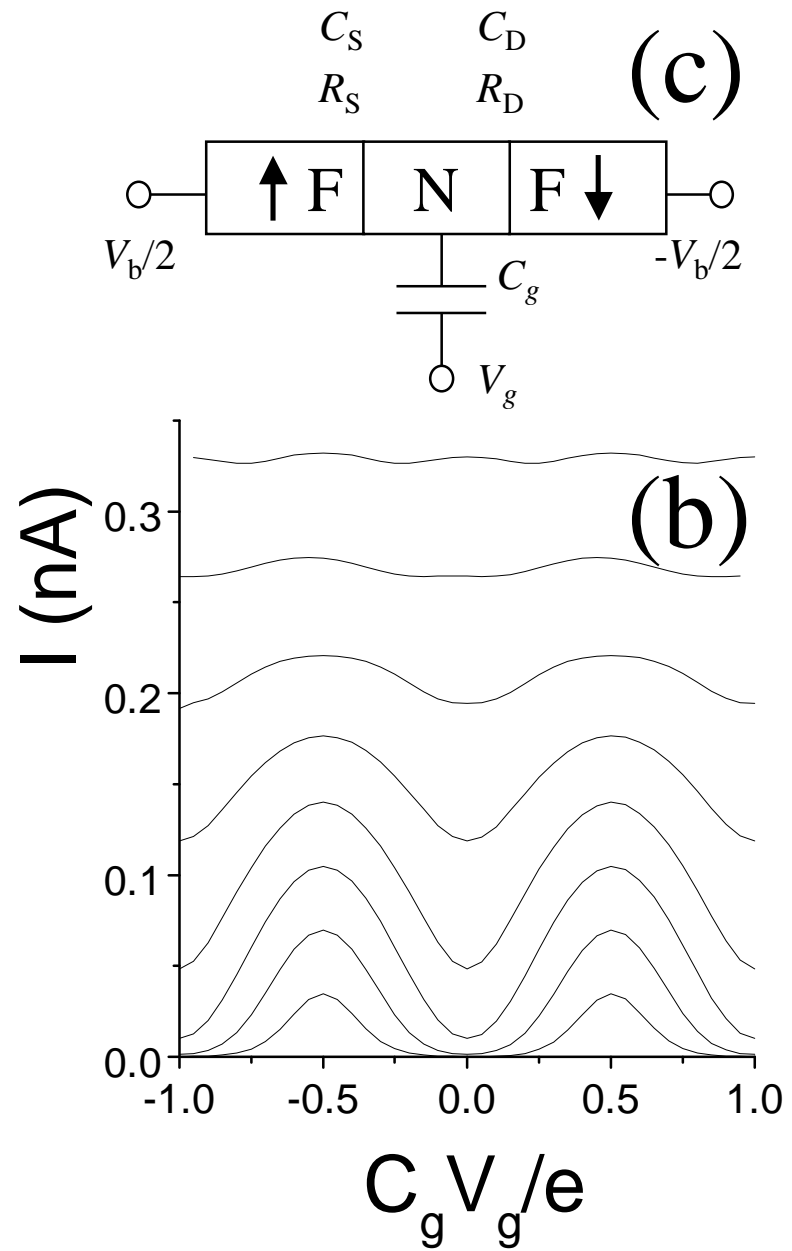
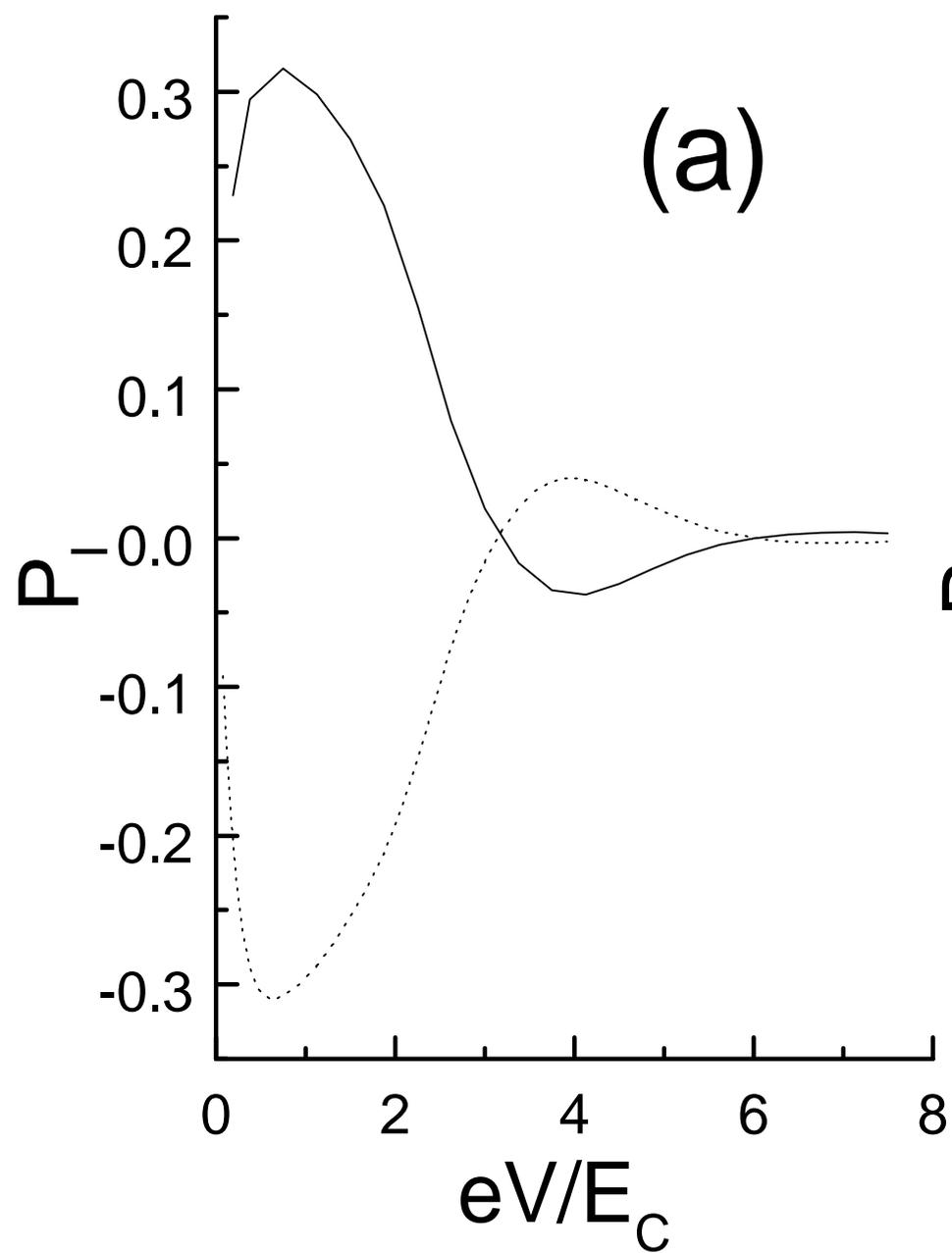
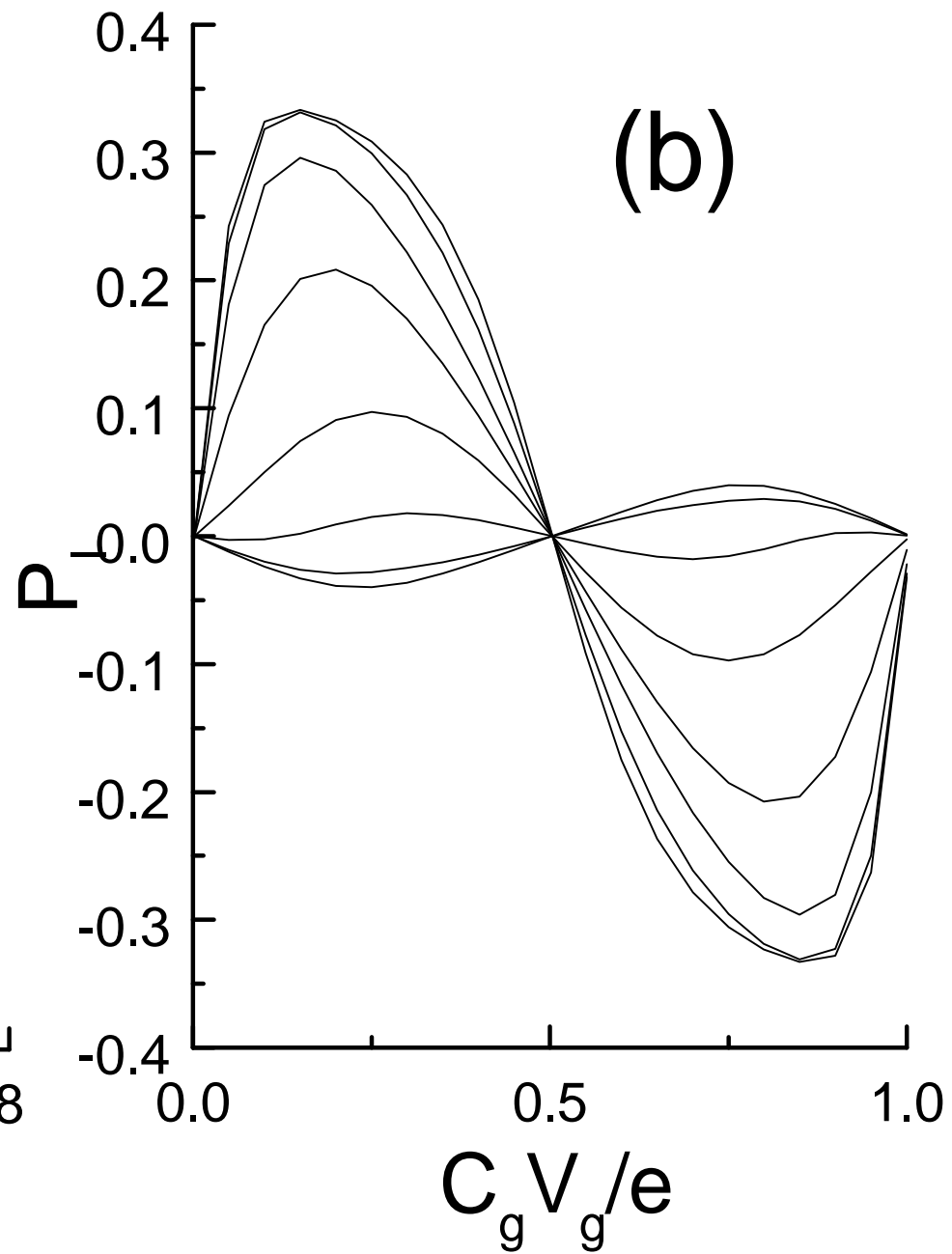
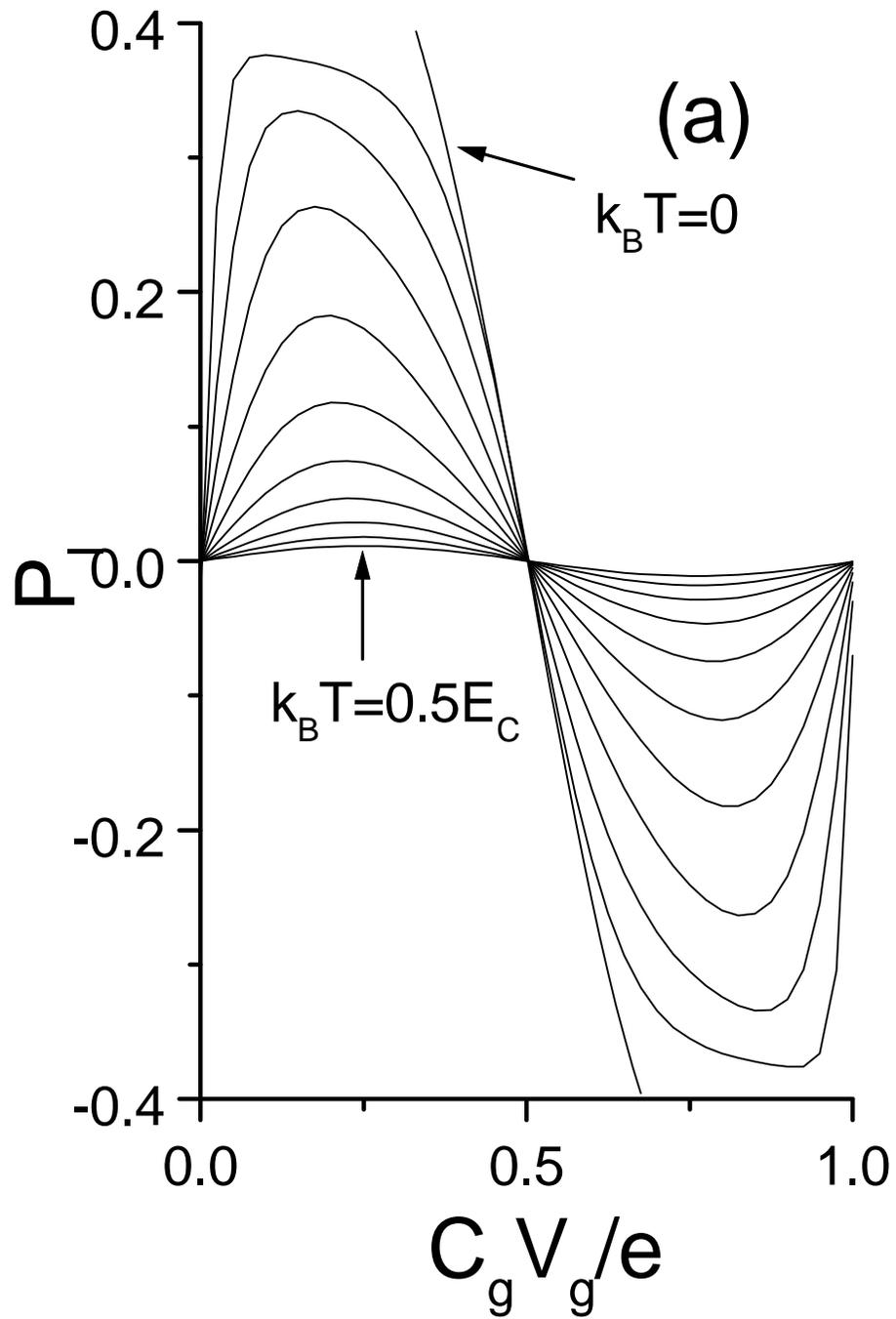
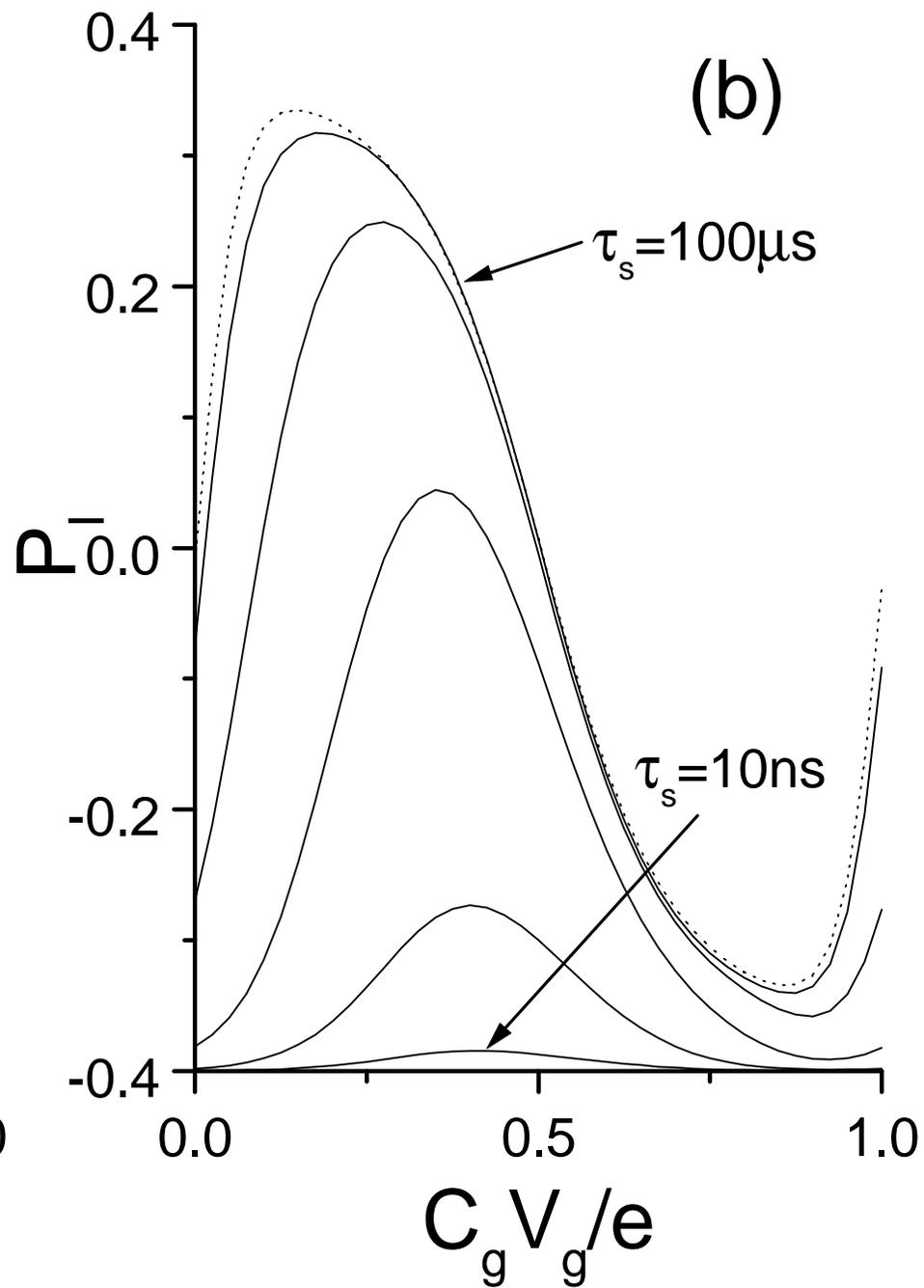
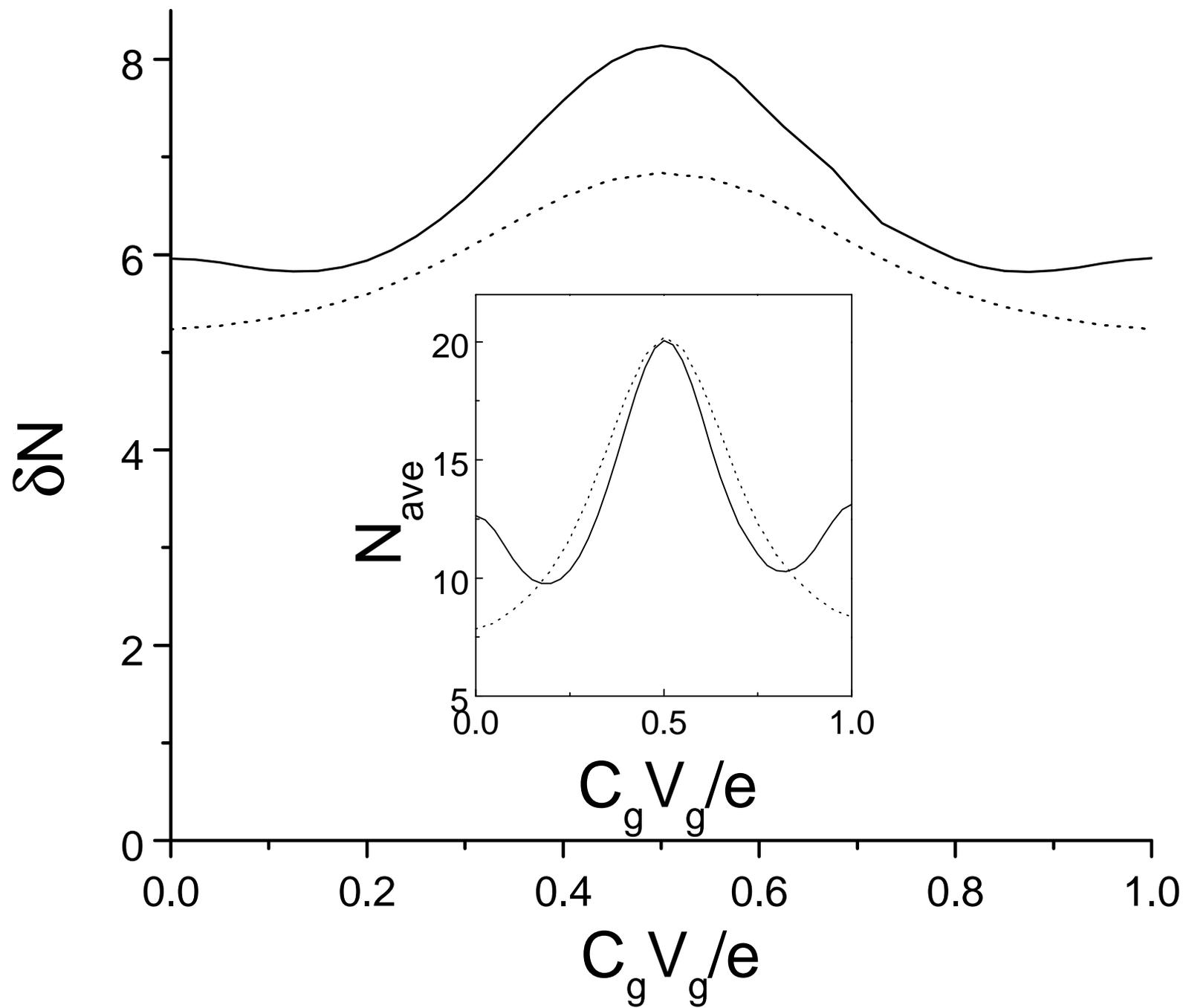
\end{document}